\documentclass[aps,prb,twocolumn,showpacs,superscriptaddress]{revtex4-1}
\usepackage{graphicx}
\usepackage{dcolumn}
\usepackage{float}
\usepackage{color}

\begin{document}

\title{Ferrimagnetism, Antiferromagnetism and Magnetic Frustration in La$_{2-x}$Sr$_{x}$CuRuO$_{6}$ (0 $\le$ \textit{x} $\le$ 1)}

\author{P. Anil Kumar}
\affiliation{Centre for Advanced Materials, Indian Association for the Cultivation of Science, Kolkata, 700032, India}
\affiliation{Department of Engineering Sciences, Uppsala University, P.O. Box 534, SE-751 21 Uppsala, Sweden}
\author{R. Mathieu}
\affiliation{Department of Engineering Sciences, Uppsala University, P.O. Box 534, SE-751 21 Uppsala, Sweden}
\author{R. Vijayaraghavan}
\affiliation{Materials Division, School of Advanced Sciences, VIT University, Vellore - 632 014, Tamil Nadu, India}
\author{Subham Majumdar}
\affiliation{Department of Solid State Physics, Indian Association for the Cultivation of Science, Kolkata 700032 India}
\author{Olof Karis}
\affiliation{Department of Physics and Astronomy, Uppsala University, Box - 516, 75120 Uppsala, Sweden}
\author{P. Nordblad}
\affiliation{Department of Engineering Sciences, Uppsala University, P.O. Box 534, SE-751 21 Uppsala, Sweden}
\author{Biplab Sanyal}
\affiliation{Department of Physics and Astronomy, Uppsala University, Box - 516, 75120 Uppsala, Sweden}
\author{Olle Eriksson}
\affiliation{Department of Physics and Astronomy, Uppsala University, Box - 516, 75120 Uppsala, Sweden}
\author{D. D. Sarma}
\email[]{Also at Jawaharlal Nehru Centre for Advanced Scientific Research, Bangalore and Council of Scientific and Industrial Research-Network of Institutes for Solar Energy (CSIR-NISE) Electronic mail: sarma@sscu.iisc.ernet.in}
\affiliation{Centre for Advanced Materials, Indian Association for the Cultivation of Science, Kolkata, 700032, India}
\affiliation{Department of Physics and Astronomy, Uppsala University, Box - 516, 75120 Uppsala, Sweden}
\affiliation{Solid State and Structural Chemistry Unit, Indian Institute of Science, Bangalore, 560012, India}

\date{\today}

\begin{abstract}
We studied structural and magnetic properties of a series of insulating double perovskite compounds, La$_{2-x}$Sr$_{x}$CuRuO$_{6}$ (0 $\le$ \textit{x} $\le$ 1), representing doping via \textit{A}-site substitution. The end members La$_{2}$CuRuO$_{6}$ and LaSrCuRuO$_{6}$ form in monoclinic structure while the intermediate Sr doped compounds stabilise in triclinic structure. The Cu and Ru ions sit on alternate B-sites of the perovskite lattice with $\sim$15\% anti-site defects in the undoped sample while the Sr doped samples show a tendency to higher ordering at B-sites. The undoped (\textit{x} = 0) compound shows a ferrimagnetic-like behaviour at low temperatures. In surprising contrast to the usual expectation of an enhancement of ferromagnetic interaction on doping, an antiferromagnetic-like ground state is realized for all doped samples (\textit{x} $>$ 0).  Heat capacity measurements indicate the absence of any long range magnetic order in any of these compounds. The magnetic relaxation and memory effects observed in all compounds suggest glassy dynamical properties associated with magnetic disorder and frustration. We show that the observed magnetic properties are dominated by the competition between the nearest neighbour Ru -- O -- Cu 180$^\circ$ superexchange interaction and the next nearest neighbour Ru -- O -- O -- Ru 90$^\circ$ superexchange interaction as well as by the formation of anti-site defects with interchanged Cu and Ru positions. Our calculated exchange interaction parameters from first principles calculations for $x$ = 0 and $x$ = 1 support this interpretation.
\end{abstract}

\pacs{75.30.Et,75.40.Gb,74.62.Dh}

\maketitle

\section{Introduction}
Double perovskite oxides with a general formula \textit{A}$_{2}$\textit{BB}${'}$\textit{O}$_{6}$ and distinct transition metal ions at \textit{B} and \textit{B}${'}$ sites are of particular interest for their varied magnetic and electronic properties. Evidently, properties of such double perovskites are strongly altered by the nature and oxidation state of the transition metal ions. For example, replacement of Mo with W in Sr$_{2}$FeMoO$_{6}$ changes the ground state from a ferromagnetic (FM) metal\cite{Nakagawa1968806,Kobayashi1998677,Sarma20002549} to an antiferromagnetic (AF) insulator.\cite{Nakagawa1969880,Ray2001607,Kobayashi2000218} Beyond this obvious route to changing properties by choosing different transition metal ions, one can also tune the electronic and magnetic properties by varying the degree of cation disorder. In particular, magnetic properties are strongly affected by such disorder, independent of the specific nature of the disorder.\cite{Kobayashi2000218,Topwal200694419,Sarma2000465,Ray2001097204,Bardelli20052} Even though the compound Sr$_{2}$FeMoO$_{6}$ has attracted huge attention because of its substantial magnetoresistance at relatively high temperatures and low magnetic fields, a wide variety of double perovskite compounds with different transition metal ions at the \textit{B} and \textit{B}${'}$ sites have been studied\cite{Philipp2003144431,Dass2004094416,Ramesha2000559,Yoshii2003236,Battle19951003} to explore diverse physical properties with an emphasis on magnetic interactions between the transition metal ions, leading to a great variety of magnetic phases within a single structural type. For example, ordered La$_{2}$\textit{M}RuO$_{6}$ is ferromagnetic for \textit{M} = Mn, is a spinglass system for \textit{M} = Fe and shows antiferromagnetic behaviour for \textit{M} = Ni and Co.\cite{Dass2004094416} The \textit{M} = Cu compound has also been investigated to probe the effect of cation ordering over \textit{B}/\textit{B}${'}$ sites on the crystal structure. It has been reported that La$_{2}$CuRuO$_{6}$ stabilizes in the monoclinic phase with a partial ordering\cite{Battle19951003} of Cu and Ru ions at \textit{B}/\textit{B}${'}$ sites. With a 50\% doping of Sr at La sites, LaSrCuRuO$_{6}$ is reported to be monoclinic with a high degree of ordering of cations\cite{Gateshki20031893} and orthorhombic with a disordered arrangement of cations.\cite{Attfield1992344} The magnetic nature of the undoped compound La$_{2}$CuRuO$_{6}$ is not established yet, while the compound LaSrCuRuO$_{6}$ has been shown to be a spinglass,\cite{Kim1993273} though the origin of the magnetic frustration responsible for the spinglass behavior has not been elaborated so far. La$_{2-x}$Sr$_{x}$CuRuO$_{6}$ is known to be insulating for 0 $\le$ \textit{x} $\le$ 1.\cite{Attfield1992344,ManjuUnpublished} In the present work, we investigate the evolution of the magnetic interactions between these two extreme compositions by studying in detail the family of compounds  La$_{2-x}$Sr$_{x}$CuRuO$_{6}$ for several values of x between 0 and 1, thereby systematically changing the formal valence state of Ru from +4 for \textit{x} = 0 to +5 for \textit{x} = 1.

\section{Experimental and Theoretical Details}
The La$_{2-x}$Sr$_{x}$CuRuO$_{6}$ compounds are prepared by the solid state reaction method. Required quantities of high purity SrCO$_{3}$, CuO, RuO$_{2}$ and La$_{2}$O$_{3}$ were mixed thoroughly. The mixture is then initially heated at 600 $^\circ$C for 24 h to avoid Ru evaporation, followed by heat treatment at 1000 $^\circ$C for 24 h and finally at 1200 $^\circ$C for 36 h. The samples are thoroughly ground before each heat treatment. The final product is then cast into pellets which are sintered at 1200 $^\circ$C for 36 h and cooled to room temperature in air. The phase purity of the product is confirmed using X-ray diffraction (XRD). The XRD patterns are recorded with a Siemens D5000 diffractometer with Cu K$\alpha$ radiation. The oxygen content of these samples as estimated from iodometric titrations is close to 6.0 for all the compositions with an accuracy of $\pm$0.05. \par A Quantum Design MPMS is used for measuring magnetization and ac susceptibility. The zero-field-cooled (ZFC) and field-cooled (FC) magnetization as a function of the temperature are  measured in the temperature range 2 - 300 K, with the measuring field set to 1 mT. The magnetization data above 100 K is used for the Curie-Weiss analysis. Magnetic memory experiments are performed according to the following protocol: the sample is cooled from 60 K to 10 K in zero field, including a stop at an intermediate temperature \textit{T}$_{S}$ for \textit{t}$_{S}$ seconds. At 10 K, a magnetic field of 1 mT is applied and the magnetic moment is measured as a function of the temperature on the heating cycle. A regular ZFC \textit{M} - \textit{T} measurement without the specific waiting protocol at an intermediate temperature is used as a reference. The magnetic relaxation data are  collected by recording the ac-susceptibility (in-phase \textit{$\chi{'}$} and out-of-phase \textit{$\chi{''}$} components with an excitation frequency of 1.7 Hz) as a function of time at 12 K after a rapid cooling from 60 K. Heat capacities of these samples are obtained using the relaxation method on a Quantum Design PPMS in the temperature range 2 - 40 K. \par We have performed first principles calculations for the two end-members, namely La$_{2}$CuRuO$_{6}$ and LaSrCuRuO$_{6}$, using VASP\cite{Kresse1993558,Kresse199611169} based on density functional theory (DFT) in the generalized gradient approximation to interpret the microscopic origin of magnetic properties observed for these compounds. A 500 eV kinetic energy cut-off was considered in the Projector Augmented Wave method. To include strong electron-electron interaction effects, we have used DFT+U approach in the Hubbard formalism. The Coulomb parameter \textit{U} and the exchange parameter \textit{J} are fixed respectively as 10 eV and 1 eV for Cu-\textit{d} orbitals while they are respectively 1.2 eV and 0.2 eV for Ru-\textit{d} orbitals. We varied the values of \textit{U} and the above-mentioned values yielded the correct magnetic states for both systems considered in the calculations. In all calculations, volume, shape and atomic positions were optimized. From the total energy calculations, the proper magnetic ground states (ferrimagnetic for La$_{2}$CuRuO$_{6}$ and antiferromagnetic for LaSrCuRuO$_{6}$) have been found.

\section{Results}
\begin{figure}
\includegraphics{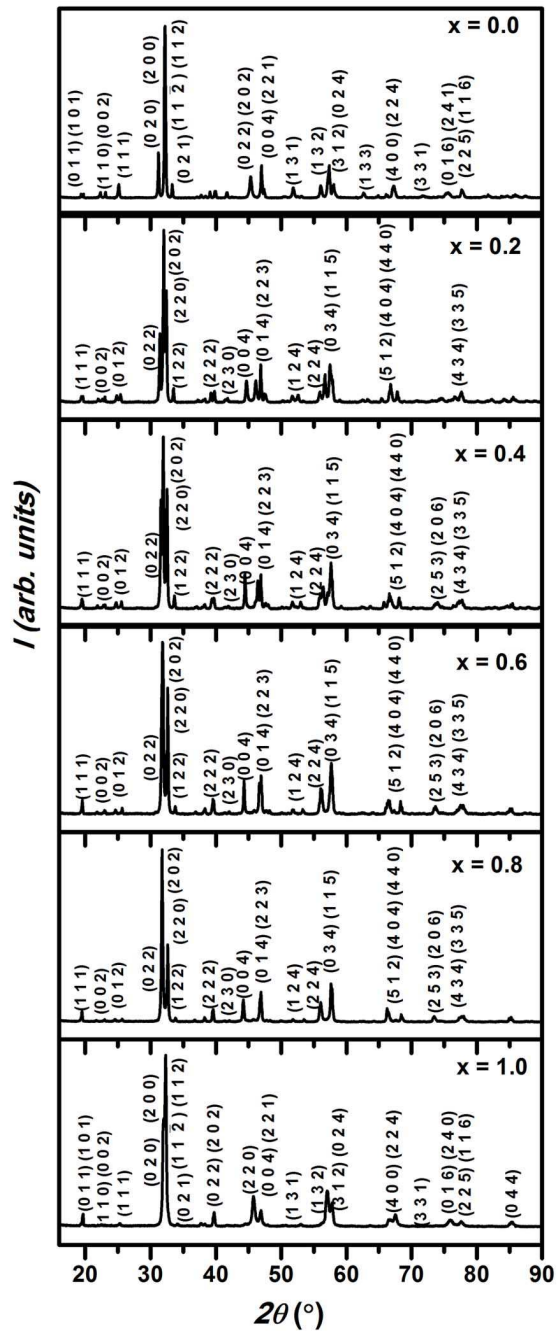}
\caption{\label{Fig.1 } The XRD patterns of the series La$_{2-x}$Sr$_x$CuRuO$_{6}$ with $x$ as indicated in each panel. The peaks are labelled with the corresponding $hkl$ indices obtained from Rietveld refinement.}
\end{figure}

\begin{table*}[t]
\caption{\label{tab:table1}%
The spacegroup symbol and lattice parameters for La$_{2-x}$Sr$_x$CuRuO$_{6}$, that are obtained from the Rietveld refinement, are presented in this table.
}
\begin{ruledtabular}
\begin{tabular}{cccccccc}
\textrm{Sr doping ($x$)}&
\textrm{Space group}&
\textrm{a}&
\textrm{b}&
\textrm{c}&
\textrm{$\alpha$}&
\textrm{$\beta$}&
\textrm{$\gamma$}\\
\colrule
0.0 & P 21/n & 5.5817 (3) & 5.7487 (3) & 7.7444 (4) & 90 & 89.87 (1) & 90\\ 
0.2 & P -1 & 7.7599 (5) & 7.8864 (5) & 8.1113 (5) & 88.921 (4) & 89.812 (7) & 90.017 (10)\\
0.4 & P -1 & 7.7560 (4) & 7.8302 (5) & 8.1512 (5) & 89.358 (4) & 89.767 (5) & 90.015 (8)\\
0.6 & P -1 & 7.7469 (5) & 7.7901 (5) & 8.1814 (5) & 89.718 (5) & 89.745 (5) & 89.982 (8)\\
0.8 & P -1 & 7.7423 (4) & 7.7656 (4) & 8.1932 (4) & 89.998 (13) & 89.765 (4) & 90.006 (18)\\
1.0 & P 21/n & 5.6127 (20) & 5.6144 (20) & 7.7680 (5) & 90 & 90.008 (35) & 90\\ 
\end{tabular}
\end{ruledtabular}
\end{table*}

\begin{figure}
\includegraphics{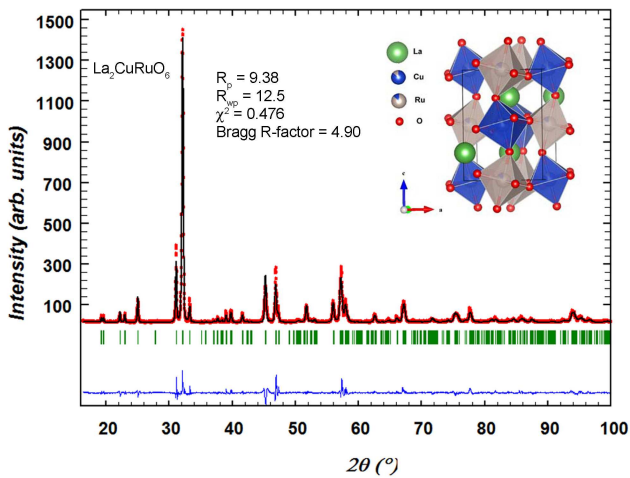}
\caption{\label{Fig.2 } (Color online) The result of the Rietveld refinement on the parent compound La$_2$CuRuO$_6$ is presented. The statistical parameters useful in judging the refinement are also mentioned. The inset shows the three dimentional image of the resultant crystal structure.}
\end{figure}

In Figure 1, we show the $hkl$-indexed XRD patterns for the series La$_{2-x}$Sr$_x$CuRuO$_{6}$ with $x$ as indicated. Rietveld refinement of these XRD data indicate that the end members, La$_{2}$CuRuO$_{6}$ and LaSrCuRuO$_{6}$, form in monoclinic structure whereas the intermediate Sr doped compounds have a triclinic crystal structure with rather small monoclinic or triclinic distortions in the lattice. We show the results of the refinement\cite{Anil2012xx} for the parent compound in Figure 2 and the three dimentional image of the resultant crystal structure, using VESTA\cite{Momma2008653}, is depicted in the inset. Table I gives the spacegroup symbol and the lattice parameters for all the compounds studied. The Reitveld refinement puts the ordering of Cu/Ru ions at \textit{B}/\textit{B}${'}$ sites in the range 90-95\% meaning an anti-site defect fraction of 5-10\% in all Sr doped samples, including LaSrCuRuO$_{6}$. However, the anti-site defect fraction in the parent La$_{2}$CuRuO$_{6}$ sample is found to be about 15\%. \par Both the undoped and doped samples are electrically highly insulating (data not shown) and their resistivity behavior is well explained by variable range hopping (VRH) conductivity. Figure 3 shows representative ZFC and FC magnetization versus temperature (\textit{M} - \textit{T}) curves for the samples \textit{x} = 0 (panel a), 0.2 (panel b) and 0.8 (panel c). In each case a clear magnetic anomaly is observed at a well defined temperature which suggests a magnetic phase transition. However, the nature of the anomaly in doped and undoped compounds differ suggesting a different type of magnetic ordering between the doped and undoped compounds. In addition, magnetic moments of doped compounds are several orders of magnitude lower compared to that of the undoped sample.
\begin{figure}
\includegraphics{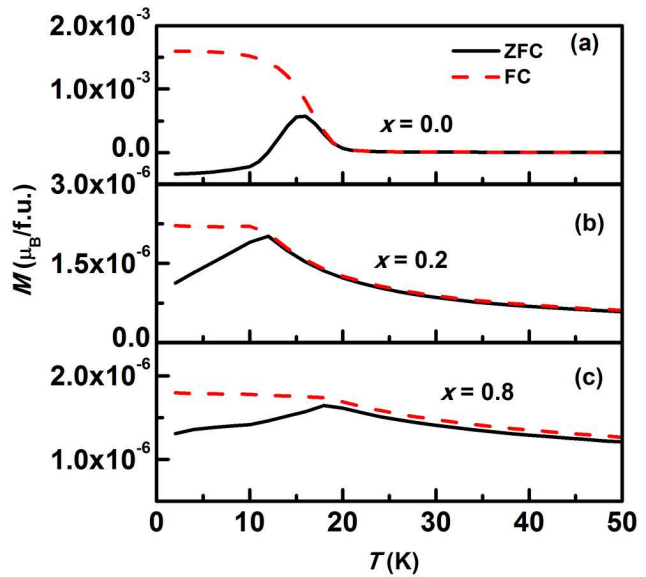}
\caption{\label{Fig.3 } (Color online) Zero field cooled and field cooled \textit{M} - \textit{T} curves shown for the La$_{2-x}$Sr$_x$CuRuO$_{6}$ samples with \textbf{(a)} \textit{x} = 0, \textbf{(b)} \textit{x} = 0.2 and \textbf{(c)} \textit{x} = 0.8.}
\end{figure}
\par The high temperature magnetization data follows a Curie-Weiss behaviour for all the samples, as shown in Figure 4(a), as manifested by the linearity of $\chi^{-1}$ vs \textit{T} plots. The values of the paramagnetic Curie temperature, $\theta_{P}$ (K) obtained from the Curie-Weiss fitting and the ZFC peak temperature, \textit{T}$_{P}$ (K) are plotted as a function of Sr doping `\textit{x}' in Figure 4(b). It is evident that $\theta_{P}$ is several times larger than the actual ordering temperature, \textit{T}$_{P}$ in the case of doped samples. $\theta_{P}$ is largely determined by the magnetic interaction strength between magnetic ions in the paramagnetic state, while the ordering temperature is in addition influenced strongly by the presence of any frustration in the magnetic interactions. Thus, the magnetic frustration parameter in such systems can be defined by the ratio of $\theta_{P}$/\textit{T}$_{P}$  which is also plotted in Figure 4(b). A negative $\theta_{P}$ for all the samples is indicative of a dominant AF interaction. The magnitude of $\theta_{P}$ increases with increasing Sr doping and so is the effective magnetic moment (\textit{p$_\mathrm{eff}$}) per formula unit which increases from $\sim$ 3.4 $\mu$$_{B}$ for \textit{x} = 0 to $\sim$ 3.8-4.0 $\mu$$_{B}$ for \textit{x} $\geq$ 0.4. The increase in \textit{p$_\mathrm{eff}$} with increasing Sr doping can be attributed to the selective oxidation of Ru from +4 to +5 with increase in Sr doping, which results in the increase of Ru spin value from 1 to 3/2 corresponding to \textit{d}$^4$ and \textit{d}$^3$ configurations, respectively. However, the increase in \textit{p$_\mathrm{eff}$} is only about 15\%, whereas $\theta_{P}$ exhibits a five fold increase across the series. This suggests that Sr doping substantially increases the AF interaction in these samples.
\begin{figure}
\includegraphics{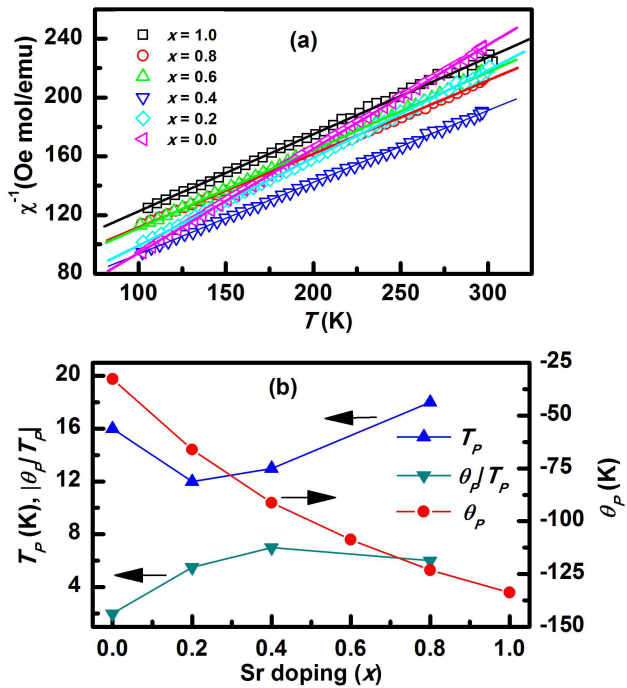}
\caption{\label{Fig.4 } (Color online) \textbf{(a)} Curie-Weiss fitting is demonstrated for the La$_{2-x}$Sr$_x$CuRuO$_{6}$ samples in the temperature range 100 to 300 K. \textbf{(b)} The characteristic temperature $\theta_{P}$, obtained from Curie-Weiss fitting is compared to the experimental transition temperature, \textit{T}$_{P}$ and the degree of frustration given by the ratio, $\theta_{P}$/\textit{T}$_{P}$ is also plotted as a function of Sr doping, \textit{x}.}
\end{figure}
\par Figure 5 shows the isothermal magnetization versus field loops for the samples with \textit{x} = 0, 0.2 and 0.8 at 5 K and 50 K. The undoped sample, \textit{x} = 0 shows a hysteresis with a coercive field, \textit{H}$_{C}\sim$ 60 mT at 5 K; while the high field magnetic moment does not quite saturate till the highest applied field accessible to us (5 T), the moment of $\sim$ 0.6 $\mu$$_{B}$ per formula unit at this highest applied field is close to the saturation moment of 1 $\mu$$_{B}$ per formula unit that is expected for an antiparallel alignment of Ru (4\textit{d}$^{4}$; $m_{S}$ = 2 $\mu$$_{B}$) and Cu (3\textit{d}$^{9}$; $m_{S}$ = 1 $\mu$$_{B}$) moments. At 50 K, the hysteresis is absent and the response is paramagnetic, consistent with the \textit{M}(\textit{T}) data in Figure 3(a) showing a magnetic transition at about 16 K. This leads to the conclusion that the \textit{x} = 0 sample is a ferrimagnet at low temperatures as suggested by the moment obtained in 5 T and the shape of \textit{M} - \textit{T} curves. The Sr doped samples do not show any hysteresis even at 5 K, and the linear \textit{M} - \textit{H} plot is basically paramagnetic both above and below \textit{T}$_{P}$ reminiscent of antiferromagnetic character. In striking contrast to the \textit{M}(\textit{T}) characteristics of the undoped (\textit{x} = 0) sample, \textit{M}(\textit{T}) data for the doped samples exhibit magnetization values several orders of magnitude lower than that of  La$_{2}$CuRuO$_{6}$. Additionally, the ZFC plots exhibit sharp maxima at 12 and 18 K for \textit{x} = 0.2 and 0.8 samples respectively, unlike the smooth and broad peak for the \textit{x} = 0 sample. Thus, it is evident from the ZFC - FC curves (Figs. 3(b) and 3(c)) in combination with the \textit{M}(\textit{H}) data (Fig. 5) that the Sr doped samples are antiferromagnetic, in contrast to the ferrimagnetic behaviour of the undoped sample. We specifically note that the doping converts the ferrimagnetic state for \textit{x} = 0 into an antiferromagnetic one for all doping levels. While there are many examples where an undoped antiferromagnetic compound is rapidly converted to a ferromagnetic state on doping, the reverse (a ferromagnetic or ferrimagnetic to antiferromagnetic phase change on doping) is rare and intriguing. \par The FC and ZFC curves in Sr doped samples, continue to diverge until high temperature, the divergence being more prominent in samples with higher Sr content.  The high temperature (well above \textit{T}$_{P}$) irreversibility between the FC and ZFC curves suggests the presence of local magnetic interactions arising from local inhomogeneities, giving rise to a distribution of ordering temperatures, extending far above the global ordering temperature, \textit{T}$_{P}$, of the majority phase. In the extreme case, this may even indicate a tiny impurity phase (below the detection limit of XRD) with a considerably higher ordering temperature.
\begin{figure}
\includegraphics{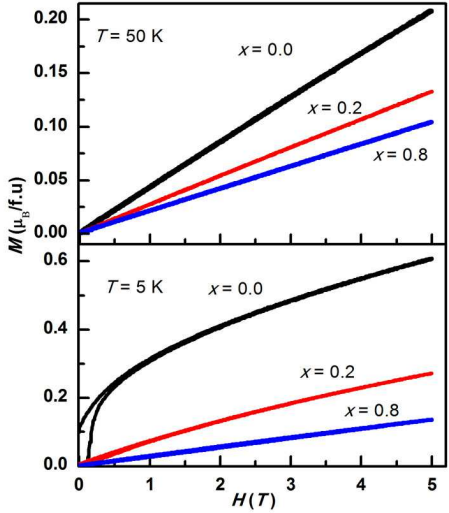}
\caption{\label{Fig.5 } (Color online) \textit{M} - \textit{H} plots at 5 K and 50 K are shown for the La$_{2-x}$Sr$_x$CuRuO$_{6}$ samples with \textit{x} = 0, 0.2 and 0.8.}
\end{figure}
\par We have further probed the dynamical magnetic properties of these compounds by performing ac magnetic measurements. Figure 6 shows the ac susceptibility data measured at three different frequencies 1.7 Hz, 17 Hz and 170 Hz for the compounds with \textit{x} = 0, 0.2, and 0.8. The external field amplitude is set to 0.4 mT. La$_{2}$CuRuO$_{6}$ shows a frequency dispersion below the magnetic anomaly and an essentially frequency independent onset of magnetic ordering. The doped samples also show some small frequency dispersion, albeit less evident than that of the undoped compound, below the maximum in $\chi$${'}$(\textit{T}), suggesting that magnetic disorder and frustration are present also in the doped compounds. One should note that the present temperature and frequency dependence of $\chi$${''}$ (or $\chi$${'}$) is quite different from those of conventional spin glasses, for which the onset of non-equilibrium dynamics ($\chi$${''}$(T)) is strongly frequency dependent.\cite{Jonsson2004174402}
\begin{figure}
\includegraphics{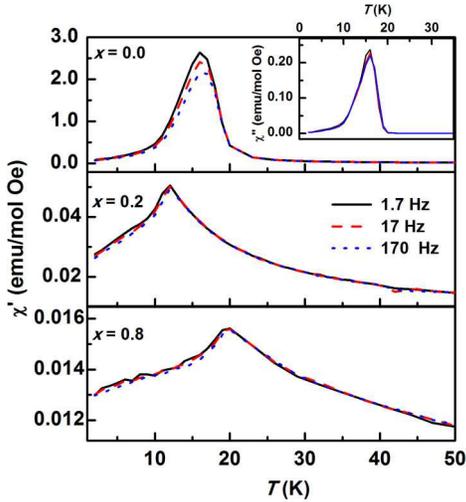}
\caption{\label{Fig.6 } (Color online) Real part of the ac susceptibility \textit{$\chi{'}$} as a function of temperature is plotted for the La$_{2-x}$Sr$_x$CuRuO$_{6}$ samples with \textit{x} = 0, 0.2 and 0.8. Imaginary part of the susceptibility is shown as inset for the sample with \textit{x} = 0.}
\end{figure}
\par Figure 7(a) shows the magnetization relaxation data, from which the glassiness in \textit{x} = 0 is evident: the out-of-phase component of the ac-susceptibility \textit{$\chi{''}$} decays with time as the spin configuration of the glassy system equilibrates or ages at constant temperature.\cite{Jonsson2004174402,Binder1986801} Figure 7(b) shows results of magnetic memory experiment which was performed according to the protocol described in Section II. The figure shows (representative data for \textit{x} = 0.8 sample) the difference between the \textit{M} - \textit{T} curves measured with the stop at intermediate temperature and the reference \textit{M} - \textit{T} curve measured without any stop. We have performed this experiment with the parameters, \textit{T}$_{S}$ = 14 K, \textit{t}$_{S}$ = 3000 seconds; \textit{T}$_{S}$ = 14 K, \textit{t}$_{S}$ = 10000 seconds; \textit{T}$_{S}$ = 16 K, \textit{t}$_{S}$ = 3000 seconds. There are clear memory dips in the curves both at 14 and 16 K, and the memory dip becomes deeper with increasing stop time (as observed for the 14 K data). The other Sr-doped samples show similar behaviour. As in the above ac-relaxation experiments, the glassy system has aged during the halt at constant temperature. This equilibration is retrieved on reheating the system, yielding the so-called memory dips seen in Fig. 7(b).\cite{Jonsson2004174402,Mathieu2001092401}
\begin{figure}
\includegraphics{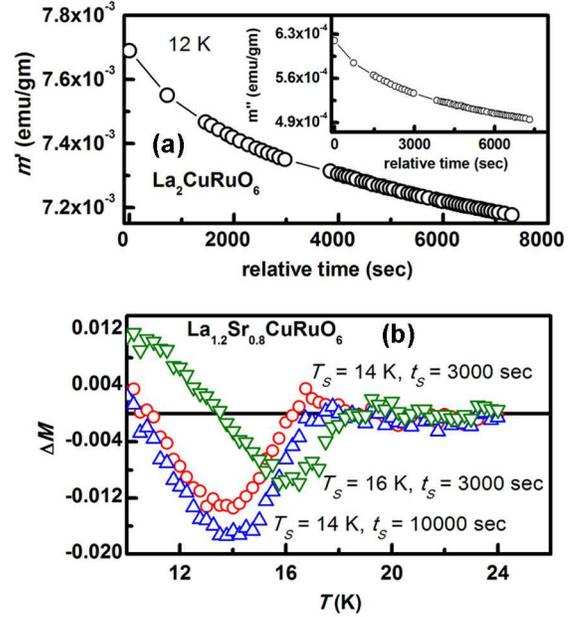}
\caption{\label{Fig.7 } (Color online) \textbf{(a)} The relaxation of real part of the magnetic moment for the \textit{x} = 0 sample (La$_{2}$CuRuO$_{6}$), the inset shows the imaginary part of the magnetization.  \textbf{(b)} The results of the memory experiments on the compound with \textit{x} = 0.8, performed according to the protocol mentioned in the main text. \textit{T}$_{S}$ and \textit{t}$_{S}$ are the stop temperature and time respectively.}
\end{figure}
\par To check for the possibility of any long range magnetic order, we have measured the heat capacity (\textit{C}) of the samples around the magnetic transition temperature. Figure 8 shows the \textit{C}/\textit{T} vs \textit{T} data for the samples with \textit{x} = 0, 0.2 and 0.4. The heat capacity data do not show any strong anomaly, but only a change of the slope for lower Sr doped samples near \textit{T$_{P}$}. Hence, the ferrimagnetic and antiferromagnetic states suggested by the magnetization measurements are not long-ranged. The \textit{C}/\textit{T} vs \textit{T} data changes with Sr doping up to \textit{x} = 0.6 and do not change significantly for \textit{x} $>$ 0.6 (see the inset of Fig. 8).
\begin{figure}
\includegraphics{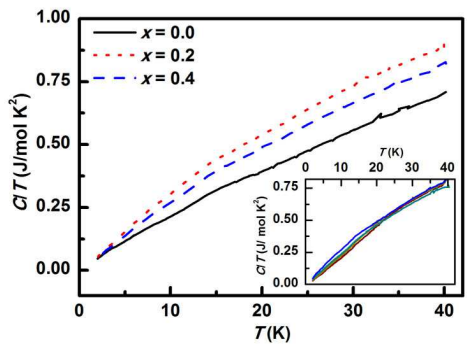}
\caption{\label{Fig.8 } (Color online) Specific heat data, shown as \textit{C}/\textit{T} vs. \textit{T} plots for La$_{2-x}$Sr$_x$CuRuO$_{6}$ samples with \textit{x} = 0, 0.2 and 0.4. The inset shows data for \textit{x} = 0.6, 0.8 and 1 along with \textit{x} = 0.4 for comparison.}
\vspace{-10pt}
\end{figure}
\section{Discussion}
The crystal sturcture of the parent compound La$_2$CuRuO$_{6}$ is shown in the inset to Figure 2. This can be treated as a representative structure of the Sr doped samples as well but with minor lattice distortions as given in Table I. It is observed that the Cu and Ru sites are octahedrally coordinated by oxygen atoms. Such corner sharing CuO$_{6}$ and RuO$_{6}$ octahedra along all three crystal axes forms the backbone of electronic and magnetic properties of this series of compounds. The La/Sr ions appearing in the space defined by the 8 nearest octahedra arranged approximately to from a distorted cube controls the electron count in the system, thereby converting Ru$^{4+}$(4\textit{d}$^{4}$) ions in the case of La$_{2}$CuRuO$_{6}$ (\textit{x} = 0) to Ru$^{5+}$(4\textit{d}$^{3}$) ions in the case of LaSrCuRuO$_{6}$ (\textit{x} = 1) compound; for intermediate values of \textit{x}, there is a disordered mixture of Ru$^{4+}$ and Ru$^{5+}$ ions in these compounds. Cu is invariably in the divalent 3\textit{d}$^{9}$ state as confirmed by XPS studies.\cite{ManjuUnpublished} 

\begin{figure}
\includegraphics{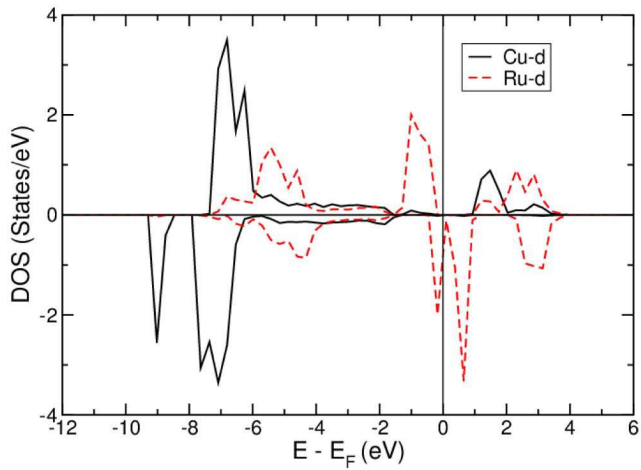}
\caption{\label{Fig.9 } (Color online) Calculated spin-polarized DOS of La$_2$CuRuO$_{6}$ from first principles. This data is for the ground state magnetic configuration mentioned in the text.}
\vspace{-10pt}
\end{figure}

The calculated partial density of states (PDOS) of La$_2$CuRuO$_{6}$, for the Cu and Ru $d$-states, are shown in Figure 9. The exchange splitting is large on the Ru atom, with a ferromagnetic coupling between the next nearest neighbour (NNN) Ru atoms. The calculated magnetic moment is 1.39 $\mu$$_{B}$ per Ru atom. For the Cu atoms the exchange splitting is smaller, with a resulting moment of 0.82 $\mu$$_{B}$ per atom. Also here the NNN interatomic coupling is ferromagnetic between the Cu atoms. As expected, the nearest neighbour (NN) Cu -- Ru exchange coupling is antiferromagnetic. The total calculated magnetic moment of the unit cell, with 4 La, 2 Cu, 2 Ru and 12 O atoms, is 2 $\mu$$_{B}$ which is mostly contributed by the Cu and Ru $d$ orbitals. Note that the additional splitting of Cu-$d$ states around 9 eV below the Fermi level occurs due to a smaller out-of-plane Cu-O bond length (calculated bond length of 1.98 \AA) compared to the other two pairs (2.12 and 2.22 \AA). Due to this, a strong hybridization occurs between Cu-$d_{z^{2}}$ and O-\textit{p$_z$} orbitals and this gives rise to a larger bonding-antibonding splitting.

\par In Figure 10 we show schematic pictures of orbitals in the \textit{xy}-plane considering only the Ru 4\textit{d}, Cu 3\textit{d} and O 2\textit{p} orbitals that are relevant for electronic, and therefore, magnetic interactions in these materials. In an ideal cubic structure, Ru and Cu alternate along the a, b and c axes with the Ru(Cu) -- O -- Cu(Ru) bond angle being 180$^\circ$, as shown in Figure 10(a). In the real crystal structure, however, of these compounds the metal-oxygen octahedra are substantially rotated (inset to Figure 2), giving rise to bond angles around 155$^\circ$ as represented schematically by off axis positioning of the O \textit{p} orbitals in Figures 10(b) and 10(c). This deviation from the linear \textit{M} -- O -- \textit{M} bonds leads to mixing of \textit{e}$_{g}$ and \textit{t}$_{2g}$ orbitals, such that both Ru \textit{t}$_{2g}$ and Cu \textit{e}$_{g}$ orbitals now can hybridize with both \textit{p$_x$} and \textit{p$_y$} orbitals at each site, thereby leading to a magnetic coupling between the Ru and Cu sites in the near neighbor positions. Inevitable presence of disorder in transition metal oxide systems is known\cite{Sarma1996307,Sarma19984004} to affect electronic structures significantly. In the present oxide family, the most significant defect is the anti-site defect, where a pair of Cu and Ru ions interchange their positions, as illustrated in the middle section of Figure 10(c), giving rise to Ru -- O -- Ru and Cu -- O -- Cu NN interactions. We show in the following that the observed magnetic properties, described in Section III, of these compounds can be accounted for by considering the interaction path-ways and the competition between the nearest neighbour (NN) and the next nearest neighbour (NNN) magnetic interactions. The possible NN interactions are between Cu and Ru  in an ordered structure via the O ion that is shared by both the Cu and Ru octahedra; anti-site defects introduce Cu-Cu and Ru-Ru interactions as well. In the following we argue that the interactions between next nearest neighbours of Ru -- O -- O -- Ru type of the ordered structure are also crucial to understand our magnetic data.

\begin{figure}
\includegraphics{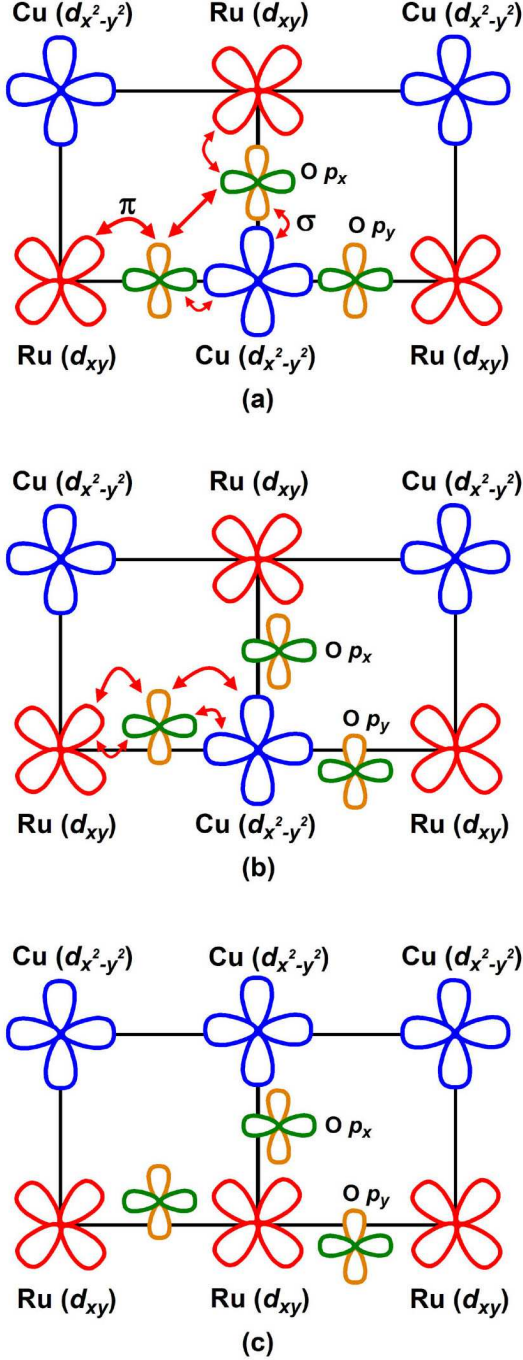}
\caption{\label{Fig.10 } (Color online) The schematic of the arrangement of O-2\textit{p}, Cu-3\textit{d} and Ru-4\textit{d} orbitals are shown and the possible nearest neighbour and next nearest neighbour hopping paths are depicted for an ideal cubic perovskite structure \textbf{(a)} and for the actual crystal structure where \textit{M} -- O -- \textit{M} bond angle is less than 180$^\circ$ \textbf{(b)}. Panel \textbf{(c)} shows the case of a anti-site defect with interchanged Cu - Ru pair (middle column) compared to \textbf(b), this anti-site defect leads to Ru - Ru and Cu - Cu NN interactions.}
\end{figure}

\subsection{\textit{x} $=$ 0}
We begin by considering various microscopic magnetic interactions in La$_{2}$CuRuO$_{6}$ (Fig. 10) where the Cu is in +2 oxidation state with the 3\textit{d}$^{9}$ (\textit{t}$^6_{2g}$\textit{e}$^3_{g}$) electronic configuration, while Ru is in the +4 oxidation state with a low spin 4\textit{d}$^{4}$(\textit{t}$^4_{2g}$) electronic configuration. The bonding between the Cu 3\textit{d} and O 2\textit{p} orbitals in the CuO$_{6}$ local octahedron is predominantly via the 3\textit{d} \textit{e}$_{g}$ and 2p$_\sigma$ hopping interactions, while that between the Ru 4\textit{d} and O 2\textit{p} in the local RuO$_{6}$ octahedron is due to 4\textit{d} \textit{t}$_{2g}$ and 2\textit{p}$_\pi$ hopping interactions. This implies that the magnetic interaction between Cu and Ru via O for an idealized cubic structure (Fig. 10(a)) with a 180$^\circ$ angle of Cu -- O -- Ru bonds will be weak, since oxygen 2\textit{p}$_\sigma$ and 2\textit{p}$_\pi$ orbitals have no on-site hopping interactions connecting them. However, in the real crystal structure, the bond angle is less than 180$^\circ$ (Fig. 10(b)) and hence the hopping between the Cu and Ru orbitals via O \textit{p} orbitals becomes finite. In this situation, the nature of the magnetic interaction depends crucially on the positioning of Cu \textit{e}$_{g}$ level with respect to Ru \textit{t}$_{2g}$ level as explained in Fig. 11(a). This figure shows schematically two possible scenarios for relative energetics of Ru and Cu \textit{d} states, one with the ferromagnetic coupling of Ru$^{4+}$ and the Cu$^{2+}$ (shown on left) and the other with the antiferromagnetic coupling of Cu$^{2+}$ and the Ru$^{4+}$ (shown on right). In these schematics in Fig. 11(a), the Cu \textit{e}$_{g\uparrow}$ and \textit{e}$_{g\downarrow}$ states are shown with an exchange splitting. In absence of any hopping interaction with Cu \textit{e}$_{g}$ states, the Ru \textit{t}$_{2g\uparrow}$ and \textit{t}$_{2g\downarrow}$ states are shown with a negligible exchange splitting in view of smaller intraatomic Hund's coupling strength of Ru 4\textit{d} states. The right-hand schematic figure shows a situation where the Ru \textit{t}$_{2g}$ states are located in the exchange gap of Cu \textit{e}$_{g}$ states. Spin conserving hopping interactions in this case splits Ru \textit{t}$_{2g}$ spin states in the way shown, following the mechanism proposed\cite{Sarma20002549} for Sr$_{2}$FeMoO$_{6}$ and shown to be valid for a large number of compounds.\cite{Saha-Dasgupta2006087205,Mahadevan2004177201} Calculated spin-polrized DOS (Fig. 9) indeed suggests this to be the case for the present compound and exhibits a spin splitting of the Ru 4$d$ levels close to 1 eV, several times larger than the Hund's coupling strength of 0.2 eV. This gives rise to an antiferromagnetic coupling between Cu and Ru \textit{d} states, as shown on the right side. In contrast, the same mechanism leads to a ferromagnetic coupling between Cu and Ru \textit{d} states with a positioning of Ru \textit{t}$_{2g}$ states outside of the exchange gap of Cu \textit{e}$_{g}$ states as illustrated in the schematic on the left-hand side of the same Fig. 11(a). The observation that the \textit{x} = 0 sample has ferrimagnetic behaviour suggests that the antiferromagnetic Cu -- O -- Ru interaction, illustrated on the right side of Fig. 11(a), is favoured compared to the ferromagnetic interaction. This is indeed confirmed by the first principles calculations (Fig. 9 and associated discussion), that shows a weak antiferromagnetic exchange coupling (J = +0.7 meV) between Cu and Ru. The weakness of the coupling strength is primarily a reflection of a small hopping strength coupling Cu \textit{e}$_{g}$ and Ru \textit{t}$_{2g}$ states arising mainly due to a deviation of Cu -- O -- Ru bond angle ($\sim$160$^\circ$) from 180$^\circ$.

\begin{figure}
\includegraphics{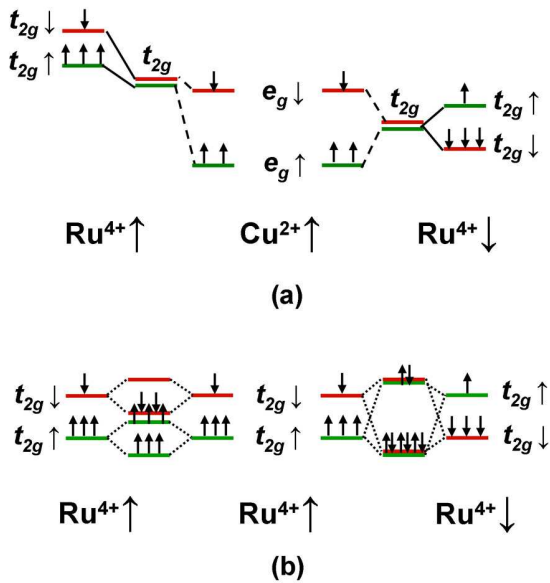}
\caption{\label{Fig.11 } (Color online) \textbf{(a)} The effect of Cu - Ru interaction on Ru energy levels is depicted in these energy level diagrams. The positioning of Ru \textit{t$_{2g}$} levels with respect to the Cu \textit{e$_{g}$} level determines the nature of Cu - Ru magnetic coupling. \textbf{(b)}  This energy level diagram shows that the Ru - Ru hopping interaction favors ferromagnetic coupling over antiferromagnetic coupling, when there is a sizable exchange splitting between Ru \textit{t$_{2g\uparrow}$} and \textit{t$_{2g\downarrow}$} states compared to the hopping strength.}
\end{figure}

\par The next nearest neighbour interaction Cu -- O -- O -- Cu is expected to be weaker, since O -- O (or O -- Ru -- O) hopping involves primarily the \textit{p}$_\pi$ orbitals, while Cu \textit{e}$_{g}$ - O \textit{p} hopping interactions involve dominantly \textit{p}$_\sigma$ orbitals. This mismatch of the orbital symmetry ensures a weak Cu -- Cu next nearest neighbour coupling; this is borne out by the calculated estimate of the ferromagnetic exchange coupling strength of -0.4 meV. Ru -- O -- O -- Ru NNN interaction will lead to a non-magnetic solution in absence of any exchange splitting of Ru \textit{d} states. However, our first principles calculations clearly show that Ru $d$ states are considerably exchange split. Therefore, the present situation with a sizeable exchange splitting of Ru \textit{t}$_{2g\uparrow}$ and \textit{t}$_{2g\downarrow}$ states compared to the hopping strength can be represented by the simplified electronic energy levels shown for the two central Ru$^{4+}$ in the schematic of Fig. 11(b). Electronic levels arising from NNN interactions for a ferromagnetic arrangement (left schematic in Fig. 11(b)) and for an antiferromagnetic arrangement (schematic on the right of Fig. 11(b)) of the two Ru$^{4+}$ are also shown. From a simple consideration of the energy level diagram, it follows that the ferromagnetic coupling leads to the lower energy state in the limit of the exchange splitting being larger than the hopping strength. This indeed is confirmed by our first principle calculations, that show the Ru 4$d$ bandwidths to be smaller than the exchange splitting, leading to a ferromagnetic ground state of the system and yielding an estimate of the ferromagnetic coupling strength of -0.8 meV for NNN interaction of Ru$^{4+}$. Therefore, the undoped ordered La$_{2}$CuRuO$_{6}$ is a ferrimagnet with antiferromagnetically coupled sublattices of ferromagnetic Cu and Ru as suggested by the NN Cu -- O -- Ru AF interactions and FM Cu -- Cu and Ru -- Ru NNN interactions. We note here that the magnetic ordering, occuring below 20 K (see Fig. 3(a)) suggests weak magnetic interactions, consistent with weak NN and NNN interactions concluded here. In addition to this overall ferrimagnetic structure, it is important to note that experimental results reported in Figs. 6 and 7 also suggest a degree of magnetic frustration in this sample. This is easily understood in terms of anti-site defects that interchange Cu and Ru positions, thereby giving rise to Ru -- O -- Ru (ferromagnetic) and Cu -- O -- Cu (antiferromagnetic) type NN interactions (represented in Fig. 10(c)). Thus, every anti-site defect leads to magnetic frustration in the otherwise perfect antiferromagnetic NN network of the ordered compound. The extent of frustration, as measured by the ratio $\theta_{P}$/\textit{T}$_{P}$, shown in Fig. 4(b) is modest for La$_{2}$CuRuO$_{6}$, suggesting that the impact of anti-site defects is relatively less for La$_{2}$CuRuO$_{6}$.

\subsection{\textit{x} $=$ 1}
Before discussing the partially doped La$_{2-x}$Sr$_{x}$CuRuO$_{6}$ compounds, it is instructive to understand the magnetic interactions present in the other end-member of this series, namely LaSrCuRuO$_{6}$. This limit is also relevant for the heavily doped compounds, such as the \textit{x} = 0.8 sample investigated here. In LaSrCuRuO$_{6}$, the Ru ion is in +5 oxidation state with a 4\textit{d}$^{3}$ electronic configuration and Cu in +2 oxidation state with the 3\textit{d}$^{9}$ electronic configuration. The nature of the Cu -- O -- Ru NN interaction can be readily understood by refering to Fig. 11(a), since no fundamental change takes place in the electronic structure and magnetic interaction in this case by the replacement of Ru$^{4+}$ \textit{d}$^{4}$ by Ru$^{5+}$ \textit{d}$^{3}$ ions. Noting that we found the Cu$^{2+}$ - Ru$^{4+}$ NN interaction to be antiferromagnetic and Ru$^{5+}$ states are likely to be more stabilized compared to that of Ru$^{4+}$, thereby helping the antiferromagnetic interaction, it is obvious that Cu$^{2+}$ - Ru$^{5+}$ NN interaction will also be antiferromagnetic. Our first principles calculations indeed supports this argument, providing an estimate of this antiferromagnetic \textit{J} = +1.1 meV, slightly larger than that (+0.7 meV) estimated for the \textit{x} = 0 compound. The major difference between \textit{x} = 0 and 1 compounds arises in the Ru -- O -- O -- Ru NNN interactions. For the \textit{x} = 1 compound, the half-filled Ru \textit{t}$^3_{2g}$ state ensures that it is a relatively strong antiferromagnetic interaction, as can be easily concluded from the electronic levels shown in Fig. 11(b) but with half-filled Ru \textit{t}$_{2g}$ levels; clearly with three \textit{t}$_{2g}$ electrons at each site, there is no stability of the ferromagnetic configuration with both the bonding and antibonding up-spin states being equally populated. Our first principles calculations estimate the Ru -- O -- O -- Ru NNN antiferromagnetic coupling strength, \textit{J}, to be +2.8 meV, thereby being the dominant interaction and controlling the magnetic state of this compound.
\begin{figure}
\includegraphics{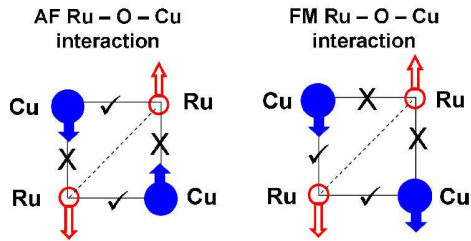}
\caption{\label{Fig.12 } (Color online) This simple diagram illustrates the magnetic frustration that is intrinsically present in the compound LaSrCuRuO$_6$. The magnetic frustration arises intrinsically irrespective of the nature of Cu - Ru interaction in the presence of a strong Ru - Ru NNN antiferromagnetic interaction. The left panel assumes antiferromagnetic coupling of Cu - Ru while the right panel assumes ferromagnetic coupling of Cu - Ru. The satisfied and unsatisfied bonds are marked with a tick and cross respectively.}
\end{figure}
Further, Fig. 12 illustrate that irrespective of Ru -- O -- Cu interaction being ferro or antiferromagnetic type, magnetic interactions in the sample are necessarily frustrated in the presence of the stronger NNN diagonal antiferromagnetic interaction. These results are therefore consistent with the spinglass behavior\cite{Kim1993273} of LaSrCuRuO$_{6}$.

\subsection{0 $<$ \textit{x} $<$ 1}
We now consider intermediate Sr doping (0 $<$ \textit{x} $<$ 1). The important distinction between these compounds and the end members (\textit{x} = 0 and \textit{x} = 1) is the existence of additional Ru(\textit{d}$^{3}$) -- O -- O -- Ru(\textit{d}$^{4}$) NNN and, in the presence of anti-site defects, Ru(\textit{d}$^{3}$) -- O -- Ru(\textit{d}$^{4}$) NN type interactions. \par The nature and the strength of Ru(\textit{d}$^{3}$) -- O -- O -- Ru(\textit{d}$^{4}$) interaction should be weakly ferromagnetic, similar to that of Ru(\textit{d}$^{4}$) -- O -- O -- Ru(\textit{d}$^{4}$) one, while that of Ru(\textit{d}$^{3}$) -- O -- O -- Ru(\textit{d}$^{3}$) is antiferromagnetic. Thus, at a modest level of doping, we would expect magnetic properties of doped La$_{2-x}$Sr$_{x}$CuRuO$_{6}$ samples to be similar to those of the undoped compound La$_{2}$CuRuO$_{6}$. This expectation would appear to be supported by the observation of similar \textit{T}$_{P}$ values for the undoped (\textit{x} = 0) and doped (\textit{x} $\ne$ 0) samples (see Fig. 4(b)). However, there is a striking distinction between the doped and undoped compounds in terms of La$_{2}$CuRuO$_{6}$ showing a ferrimagnetic-type transition, while all doped samples exhibit antiferromagnetic-type transitions. While there is a large number of examples of an antiferromagnetic undoped compound being converted to a ferromagnetic one on doping, the reverse case of a ferro or ferrimagnetic system changing over to an antiferromagnetic state is extremely rare and intriguing from a microscopic point of view. In order to understand this unusual phenomenon, we first note that anti-site defects involving Ru$^{5+}$ ions will give rise to Ru$^{4+}$ -- O -- Ru$^{5+}$ ferromagnetically coupled NN cluster, as shown in Fig. 13 in terms of the cluster of open circles representing Ru$^{4+}$ ions and an open triangle representing a Ru$^{5+}$ ion at an anti-site position. Simultaneously generated Cu$^{2+}$ -- O -- Cu$^{2+}$ will be strongly coupled antiferromagnetically due to superexchange interactions as also shown in Fig. 13. A direct consequence of this is to form two ferrimagnetic domains of  La$_{2}$CuRuO$_{6}$-like clusters on the two sides of the defect with these two domains being \textit{antiferromagnetically coupled} as illustrated in Fig. 13. Thus, the long range magnetic order in the system will be controlled by these antiferromagnetic coupling of the small ferrimagnetic domains; this is also expected to reduce the total magnetization drastically, as indeed observed in Fig. 3.
\begin{figure}
\includegraphics{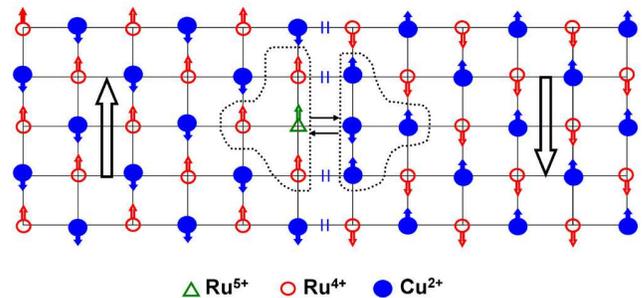}
\caption{\label{Fig.13 } (Color online) Schematic illustrates the consequence of an anti-site defect on the magnetic structure of the sample. The inter-changed Cu -- Ru pair is shown by small horizontal arrows. This inter-change leads to some frustrated Cu -- Ru NN interactions which are marked with two short (blue) vertical lines through the Cu - Ru bonds. Rest of the lattice is governed by the Cu -- O -- Ru NN antiferromagnetic interactions. As designated by the directions of the large vertical arrows, the anti-site defects leads antiferromagnetically coupled clusters of ferrimagnetic domains on two sides.}
\end{figure}
\par We note that an increased level of doping gives rise to an enhanced magnetic interaction strength, as suggested by a modest increase in the ordering temperature, \textit{T}$_{P}$, and a rapid increase in $\theta_{P}$ (see Fig. 4). We also find that frustration parameter, $\theta_{P}$/\textit{T}$_{P}$, rapidly increases with \textit{x} for small values of \textit{x}, attaining a saturation value of $\sim$ 7.0 for \textit{x} $\ge$ 0.4. Additionally, we also find that the irreversibility in terms of the separation of the FC and ZFC curves (Fig. 3) extends to a very high temperature for high values of doping. These interesting observations can be easily understood in terms of the basic interactions already discussed here. \textit{T}$_{P}$ in the doped samples are controlled by the presence of Ru$^{5+}$ sites and the associated anti-site defects; therefore, the increase in \textit{T}$_{P}$ can be associated with the increasing abundance of Ru$^{5+}$ ions with an increasing doping level in this series of compounds. In contrast to the modest increase in \textit{T}$_{P}$, $\theta_{P}$ increases by nearly an order of magnitude, indicating this to have a different origin compared to that for increase in \textit{T}$_{P}$. We note that the increase in doping leads to the formation of increasing sites with the half-filled \textit{d}$^{3}$ configuration of Ru$^{5+}$. Thus, an increasing \textit{x} gives rise to an increasing number of Ru$^{5+}$(\textit{d}$^{3}$) -- O -- O -- Ru$^{5+}$(\textit{d}$^{3}$) pairs with antiferromagnetic interaction that has already been discussed in the context of the fully doped (\textit{x} = 1) LaSrCuRuO$_{6}$ compound. In this limit, anti-site defects also give rise to antiferromagnetic Ru$^{5+}$(\textit{d}$^{3}$) -- O -- Ru$^{5+}$(\textit{d}$^{3}$) pairs. Thus, progressive doping shifts the dominant magnetic interactions in the system from being those of La$_{2}$CuRuO$_{6}$ that are weak to those of LaSrCuRuO$_{6}$ that are relatively stronger, accounting for the steady increase in $\theta_{P}$ with \textit{x}. Similarly, the origin of frustration in magnetic interactions is distinctly different at the two ends of this series. The mangetic frustration is governed by anti-site defects in the low \textit{x}-value regime, while magnetic frustration is built into the microscopic magnetic interactions of even the ordered \textit{x} = 1 sample, arising from a dominant antiferromagnetic interaction along the face diagonal of the cell. Thus, the increase in the frustration parameter is related to a change over from the weak frustration arising from anti-site defects, already discussed for \textit{x} = 0 compound, for small values of \textit{x} to a strong frustration based on the intrinsic magnetic interactions discussed in the case of \textit{x} = 1 compound. The irreversibility extending to a much higher temperature for the highly doped samples arises for this more robust frustration prevalent in the large \textit{x} limit due to the abundance of Ru$^{5+}$(\textit{d}$^{3}$) ions.

\section{Conclusions}
We have studied structural and magnetic properties of a series of double perovskite copper ruthenate compounds, La$_{2-x}$Sr$_{x}$CuRuO$_{6}$ (0 $\le$ \textit{x} $\le$ 1). While the undoped compound, La$_{2}$CuRuO$_{6}$ shows the characteristics of a short range ferrimagnet, even the smallest Sr doping (\textit{x} = 0.2) changes the ground state basically to an antiferromagnetic one with glassy dynamics. These properties can be adequately explained by considering the competing nearest neighbour and next nearest neighbour transition metal-transition metal interactions of Cu -- Ru, Ru -- Ru and Cu -- Cu types, including such pairs arising from anti-site defects. Structural distortions that cause a significant deviation of the Ru -- O -- Cu bond angle from 180$^\circ$ of the ideal cubic structure is also found to play an important role in determining the magnetic interactions. Interestingly, magnetic interaction strengths between different NNN pairs connected by the 90$^\circ$ interactions are found to be comparable to those between  NN pairs along the bond via oxygen ($\approx$ 180$^\circ$ interaction). Such a situation, not encountered for compounds of only 3$d$ transition elements, arises due to the presence of Ru ions with its more extended 4$d$ orbitals and the partial occupancy of $t_{2g}$ orbitals that interact with oxygen ions via $\pi$-interactions. We find that this substantial, and at times dominant, NNN interactions and the presence of anti-site defects are crucial to understand the most interesting properties of this series of compounds, such as the conversion of the undoped $x$=0 compound to an essentially antiferromagnetic one on doping ($x~>$~0) as well as the evidence of frustration and glassy dynamics for all values of $x$. It is found that frustration is dominated by the anti-site defects for small $x$, while the NNN antiferromagnetic Ru-Ru interaction dominates the galssy dynamics at larger values of $x$.

\begin{acknowledgments}
Authors thank the Department of Science and Technology, Government of India and Swedish Foundation for International Cooperation in Research and Higher Education (STINT) for supporting this research. PAK, RM and PN thank the Swedish Research Council (VR) and the G\"{o}ran Gustafsson Foundation, Sweden for funding. DDS thanks the national J. C. Bose Fellowship. DDS and OE gratefully acknowledge the financial support from the Swedish Research Links programme funded by VR/SIDA, Sweden. OE also thanks the KAW foundation as well as the ERC. BS thanks Marjana Lezaic of Forschungszentrum, Juelich, Germany for useful discussions. Authors thank Prof Sergey Ivanov for the expert help with the Rietveld refinement.
\end{acknowledgments}

%

\end{document}